\documentclass[reprint,eqsecnum,floats,aps,amsmath,amssymb,nofootinbib,prd,onecolumn, showpacs]{revtex4-1}
\usepackage{eufrak}
\usepackage{graphicx}
\usepackage{amsmath,amssymb}
\usepackage{hyperref}
\usepackage{graphicx}
\usepackage[utf8x]{inputenc}
\usepackage{setspace}

\begin{document}
	
\author{Beatriz Elizaga Navascu\'es}
\email{beatriz.b.elizaga@gravity.fau.de}
\affiliation{Institute for Quantum Gravity, Friedrich-Alexander University Erlangen-N{\"u}rnberg, Staudstra{\ss}e 7, 91058 Erlangen, Germany}
\author{Guillermo A. Mena Marug\'an}
\email{mena@iem.cfmac.csic.es}
\affiliation{Instituto de Estructura de la Materia, IEM-CSIC, Serrano 121, 28006 Madrid, Spain}
\author{Santiago Prado}
\email{santiago.prado@iem.cfmac.csic.es}
\affiliation{Instituto de Estructura de la Materia, IEM-CSIC, Serrano 121, 28006 Madrid, Spain}

\title{Unique fermionic vacuum in de Sitter spacetime from hybrid quantum cosmology} 

\begin{abstract} 

In this work we show how the criterion of asymptotic Hamiltonian diagonalization originated in hybrid quantum cosmology serves to pick out a unique vacuum for the Dirac field in de Sitter, in the context of quantum field theory in curved spacetimes. This criterion is based on the dynamical definition of annihilation and creationlike variables for the fermionic field, which obey the linearized dynamics of a Hamiltonian that has been diagonalized in a way that is adapted to its local spatial structure. This leads to fermionic variables that possess a  precise asymptotic expansion in the ultraviolet limit of large wavenumbers. We explicitly show that, when the cosmological background is fixed as a de Sitter solution, this expansion uniquely selects the choice of fermionic annihilation and creationlike variables for all spatial scales, and thus picks out a unique privileged Fock representation and vacuum state for the Dirac field in de Sitter. The explicit form of the basis of solutions to the Dirac equation associated with this vacuum is then computed.

\end{abstract}


\maketitle 

\newpage

\section{Introduction}

One of the solutions of General Relativity (GR) that deserves special attention in modern cosmology is de Sitter spacetime.  This is because, in the context of primordial cosmology, this solution approximates quite well  the expected behaviour of an inflationary period in the evolution of the Universe. The practical benefits of this approximation are numerous, among which it is remarkable the application to the quantum field theory (QFT) description of cosmological perturbations. Indeed, for cosmological inflationary models driven by a spin-$0$ matter field (the inflaton), the standard theoretical guideline for the choice of an initial quantum state of the inhomogeneous perturbations in the metric and the inflaton at the onset of inflation is to select an analogue of the Bunch-Davies (BD) vacuum \cite{BD,mukh}. This is, in turn, the preferred Fock vacuum of a quantum scalar field propagating in de Sitter, that is picked out as the unique Hadamard state among those that are invariant under the isometry group of de Sitter spacetime, that is maximally symmetric \cite{allen,bidav,Hadamard}. Interestingly, the power spectra of cosmological perturbations that is predicted in GR with such a choice of BD vacuum agrees, to a high degree of accuracy, with the current experimental observations of the Cosmic Microwave Background \cite{planck,planck-inf,planck1,planck2}.

The physical and mathematical properties of the BD vacuum (and its associated Fock representation) for scalar fields have been thoroughly studied in the literature (see e.g. Refs. \cite{bd1,bd2,bd3,bd4,bd5,bd6,ssasaki,bdunit}). For higher spin fields, and in particular for the case of the spin-$1/2$ Dirac field, generalizations of the notion and features of the BD vacuum have also been widely discussed in the literature \cite{bdf1,bdf2,bdf3,bdf4}, even though the uniqueness of the resulting state might not be as broadly established as for scalar fields. The aim of this paper is to justify a physically natural choice of Fock vacuum state in de Sitter spacetime, for a minimally coupled Dirac field, that displays BD-like properties and that finds its motivation in the context of quantum cosmology. More specifically, we will explicitly derive the unique Fock representation that turns out to be selected by the criterion of asymptotic Hamiltonian diagonalization, recently introduced for the so-called hybrid approach to canonical quantum cosmology \cite{diagfermi}.

Hybrid quantum cosmology \cite{hlqc1,hlqc2} is an strategy for the canonical quantization of spacetimes that contain inhomogeneities, but also possess some notion of symmetry. Such is the case, for instance, of the system formed by matter and metric perturbations over an otherwise homogeneous and isotropic inflationary cosmology of the Friedmann-Lema\^{\i}tre-Robertson-Walker (FLRW) type \cite{hybr-inf1,hybr-inf2,hybr-inf3,hybr-ref,hybr-ten,fermihlqc}. In particular, the quantization strategy is based on the use of different representation techniques for the canonical algebra that describes the homogeneous degrees of freedom, on the one hand, and for the algebra that contains the inhomogeneous fields, on the other hand. The homogeneous algebra is represented with techniques imported from a theory of quantum gravity or quantum cosmology, e.g. loop quantum gravity \cite{lqg}, while the inhomogeneous fields are given a more conventional Fock representation. From a theoretical point of view, the first important condition that one should require from this approach is that the combination of the two different quantum representations consistently leads to nicely defined operator versions of the constraints of the gravitational system. To that end, it becomes necessary to carry out a sensible choice of canonical splitting between the homogeneous and inhomogeneous sectors of phase space, in view of their posterior quantization, taking into account the qualitatively different quantum descriptions that are going to be adopted for them. 

Declaring which part of the phase space should correspond to the inhomogeneous fields in hybrid quantum cosmology has been the central point of several investigations. Actually, this ambiguity can be codified altogether with the freedom in the choice of a Fock representation for them, by means of families of variables that are the classical counterpart of the annihilation and creation operators, obtained through canonical transformations in the entire phase space that respect the basic symmetries of the homogeneous sector. The first physical criterion for any admissible Fock representation in hybrid quantum cosmology should then be that, in the context of QFT in curved spacetimes (and hence regarding the homogeneous sector as classical), the annihilation and creation operators undergo an evolution that can be implemented by a quantum unitary transformation \cite{unig1,unig2,uni1,uni2,uni3,uniqueness1,uniqueness2,uniqueness3,uniqfermi}. This criterion of choice has been further restricted in the context of fermionic perturbations in hybrid quantum cosmology, in such a way that certain effective backreaction to the Hamiltonian constraint does not develop divergences, without the need of introducing any regularization procedure  \cite{backreaction}. Finally, and motivated by these previous conditions, the most recent works on the theoretical formulation of the hybrid quantization of both fermionic and scalar cosmological perturbations have proposed an approach that aims to remove all the ambiguities (up to irrelevant phases) in the choice of the canonical algebra associated with the perturbations, that is going to be quantized \`a la Fock \cite{diagfermi,diagscalar}. The approach tries and constructs a quantum description of the inhomogeneities such that the local structure of the Hamiltonian contains no self-interaction contributions in terms of the annihilation and creation operators, and thus it is asymptotically diagonalized in the ultraviolet regime of short scales. Such a criterion turns out to completely fix, in an asymptotic expansion, the dynamical definition of those operators, expansion from which one may hope to uniquely determine them globally. In the restricted context of QFT in a de Sitter background cosmology, this hope was actually realized for the scalar perturbations, resulting in the specification of the well-known BD vacuum \cite{diagscalar}. In this paper, we show in detail how the criterion of asymptotic diagonalization also serves to uniquely fix the choice of vacuum in de Sitter for fermionic perturbations of the Dirac type. The result provides a privileged Fock representation of the Dirac field in de Sitter spacetime for which the selection criterion is univocally characterized and well understood.

The structure of this paper is as follows. In Sec. \ref{diaggeneral} we introduce the fermionic perturbations for a general FLRW cosmological background, and the fermionic variables that are going to be promoted to annihilation and creation operators that display a dynamical evolution that is dictated by a diagonal Hamiltonian. We summarize the procedure to construct these variables, that is based on an asymptotic diagonalization. In Sec. \ref{desitter} we explicitly solve the problem of finding the most general family of annihilation and creation operators that evolve without self-interaction in a fixed de Sitter background. We then show how the method of asymptotic diagonalization of the Hamiltonian, adapted to its local structure, serves to fix a unique set of annihilation and creation operators, and give the specific form of the corresponding privileged Fock representation. Finally, in Sec. \ref{conclusions} we summarize our results.

Throughout the text, we employ natural units, so that $G=c=\hbar=1$.

\section{Fermionic vacuum in hybrid quantum cosmology}\label{diaggeneral}
   
In this paper we focus our attention on a system that has been extensively studied in hybrid quantum cosmology: a spatially flat homogeneous FLRW spacetime which is minimally coupled to a homogeneous scalar field with a certain potential, as well as to an inhomogeneous Dirac field. The spatial hypersurfaces of this cosmology are taken to be compact, isomorphic to the three torus $T^{3}$. The FLRW metric can be described in terms of a scale factor $\tilde{a}$. On the other hand, we treat the Dirac field entirely as a perturbation of the system. Moreover, whenever we consider canonical transformations that mix the fermionic degrees of freedom with the homogeneous background, we adopt a truncation scheme in which we only preserve contributions to the Einstein-Dirac action that are at most quadratic in the perturbations. The transformations are viewed as canonical within the framework of this perturbative truncation, including the symplectic structure of the system. Potentially, one may also introduce perturbations of the metric and the scalar field, and work consistently within our truncation scheme and a canonical framework for the entire cosmological system \cite{hybr-ref,hybr-ten,fermihlqc}.

For the description of the Dirac field $\psi$, we use the Weyl representation of the constant Clifford algebra associated with the four-dimensional flat metric,
\begin{align}
\gamma^{0}=i
\begin{pmatrix}
0 & I \\ I & 0
\end{pmatrix},\qquad \vec{\gamma}=i
\begin{pmatrix}
0 & \vec{\sigma} \\ -\vec{\sigma} & 0
\end{pmatrix},
\end{align}
where $I$ denotes the identity matrix (here, in two dimensions), $\vec{\gamma}=(\gamma^{1},\gamma^{2},\gamma^{3})$, and $\vec{\sigma}=(\sigma_1 ,\sigma_2 ,\sigma_3)$ is the triple formed by the three Pauli matrices. In addition, we fix the gauge time direction of the tetrads so that it coincides with the (future-directed) normal vector to the homogeneous spatial hypersurfaces, in order to simplify the Dirac brackets of the field \cite{teitelboim} and so that, with respect to its local Lorentz transformation properties, we can regard it as two representation spaces of $SU(2)$, rather than $SL(2,\mathbb{C})$ \cite{uniqueness3}.  Then, we exploit the high symmetry and compactness of the flat homogeneous spatial hypersurfaces in order to decompose the Dirac field in terms of a complete set of modes \cite{uniqueness3,uniqfermi,fermihlqc},
\begin{align}\label{mdecomp}
\psi (t,\vec{x})=\sum_{\vec{k}\in\mathbb{Z}^3}\sum_{\lambda=\pm 1}\frac{e^{i2\pi\vec k\vec{x}/l_0}}{\sqrt{l_0^3 \tilde{a}^3}}\begin{pmatrix}
x_{\vec k ,\lambda}(t)\xi_{\lambda}(\vec{k}) \\ 
y_{\vec k ,\lambda}(t)\xi_{\lambda}(\vec{k})
\end{pmatrix},\qquad -i\vec{\sigma}\vec{\nabla}\left[\xi_{\lambda}(\vec{k})e^{i2\pi\vec k \vec{x}/l_0}\right]=\lambda\omega_k\xi_{\lambda}(\vec{k})e^{i2\pi\vec k\vec{x}/l_0}.
\end{align}
Here, $l_0$ is the compactification length of the torus, $\lambda\omega_k$ are the eigenvalues of the Dirac operator $-i\vec{\sigma}\vec{\nabla}$ on $T^3$, where $\lambda=\pm 1$ represents the helicity, the two-component objects $\xi_{\lambda}(\vec{k})\exp({i2\pi\vec k\vec{x}/l_0})$ are its eigenspinors, and $\omega_k=2\pi\vert \vec k\vert/l_0$ with  $\vec{k}\in \mathbb{Z}^3$. In addition, we are imposing spatially periodic boundary conditions to the Dirac field, restricting ourselves in this way to the trivial choice of spin structure in $T^3$ \cite{dtorus}. The time-dependent coefficients $(x_{\vec k ,\lambda},y_{\vec k ,\lambda})$ are Grassmann variables that only display non-vanishing Dirac brackets with their complex conjugates, each of these non-trivial brackets being equal to $-i$. 

In the truncation scheme that we have adopted, these fermionic variables contribute to the total Hamiltonian of the system only through the zero-mode of the Hamiltonian constraint. Explicitly, this fermionic contribution to the Hamiltonian is given by \cite{fermihlqc}
\begin{align}\label{hdorig}
\tilde{H}_D={N}_{0}\sum_{\vec{k},\lambda}\left[M  \big( {\bar y}_{\vec{k},\lambda} x_{\vec{k},\lambda} + {\bar{x}}_{\vec{k},\lambda} { y}_{\vec{k},\lambda}\big) 
-  \tilde{a}^{-1}\lambda \omega_k  \big( {\bar{x}}_{\vec{k},\lambda} x_{\vec{k},\lambda} - {\bar{y}}_{\vec{k},\lambda} y_{\vec{k},\lambda} \big)\right],
\end{align}
where $M$ is the bare mass of the Dirac field, ${N}_{0}$ is the lapse function of the homogeneous FLRW background, and an overbar indicates complex conjugation. We exclude from all of our considerations and sums the terms with $\vec{k}=0$, namely the fermionic zero-modes, since they contribute to the Hamiltonian in a slightly different manner and can be isolated and quantized separately. For instance, one could directly adopt a standard holomorphic representation for the (finitely many) anticommuting variables that describe these zero-modes (for details on this type of quantization, see e.g. Ref. \cite{DH}). On the other hand, let us notice that modes corresponding to different values of $\vec{k}$ and $\lambda$ completely decouple in $\tilde{H}_D$, and that the coefficients of the fermionic variables only depend on $\lambda\omega_k$, but not on the degeneracy of these eigenvalues of the Dirac operator (i.e. they only depend on the helicity and on the norm of $\vec{k}$). This is mostly a manifestation of the spatial symmetries of the spatial sections, together with the conservation of the helicity of the field in FLRW cosmologies \cite{uniqfermi}.

The freedom that exists in hybrid quantum cosmology in the way to split the phase space into a homogeneous sector and an inhomogeneous sector can be captured in the choice of a background-dependent family of variables of annihilation and creation type for the description of the dynamical Dirac field, respecting the symmetries of the Hamiltonian that we have commented in the above pragraph. These families of variables are of the general form
\begin{align}\label{ab}
 \begin{pmatrix}
a_{\vec k ,\lambda} \\ 
\bar{b}_{-\vec k ,\lambda}
\end{pmatrix} 
=
\begin{pmatrix}
f_1^{k, \lambda} & f_2^{k, \lambda} \\ g_1^{k, \lambda} & g_2^{k, \lambda}
\end{pmatrix}\left[I-\frac{1-\lambda}{2}(I-\sigma_1)\right]\begin{pmatrix}x_{\vec k ,\lambda} \\  y_{\vec k ,\lambda}
\end{pmatrix}.
\end{align}
As we have indicated, the coefficients $f_{l}^{k,\lambda}$ and $g_{l}^{k,\lambda}$, with $l=1,2$, are in principle allowed to depend on the canonical variables that determine the homogeneous cosmological background: the scale factor $\tilde{a}$, the homogeneous scalar field, and their canonical momenta. These coefficients are subject to the following relations:
\begin{align}\label{fg}
f_2^{k, \lambda}=e^{iF_2^{k, \lambda}}\sqrt{1-\left\vert f_1^{k, \lambda}\right\vert^2},\qquad g_1^{k, \lambda}=e^{iJ_{k, \lambda}}\bar f_2^{k, \lambda},\qquad g_2^{k, \lambda}=-e^{iJ_{k, \lambda}}\bar f_1^{k, \lambda},
\end{align}
that ensure that the transformation of the pair $(x_{\vec k ,\lambda},y_{\vec k ,\lambda})$ to $(a_{\vec k ,\lambda},\bar{b}_{-\vec k ,\lambda})$ is canonical with respect to the symplectic structure restricted to the fermionic sector of phase space. Here, $F_2^{k, \lambda}$ and $J_{k, \lambda}$ are unspecified phases. The standard convention is then to regard $a_{\vec k ,\lambda}$ as the pre-quantum version of annihilation operators of particles, and $\bar{b}_{\vec k ,\lambda}$ as the variables that are going to be promoted to creation operators of antiparticles. Every single specification of such variables for all wavevectors $\vec{k}$ defines a different Fock quantization of the fermionic field. 

We notice that, since the coefficients $f_{l}^{k,\lambda}$ and $g_{l}^{k,\lambda}$ that define them depend generally on the homogeneous cosmological background or, from a classical perspective, on time, the Hamiltonian that dictates the linearized classical dynamics of these annihilation and creationlike variables is different from $\tilde{H}_{D}$ in Eq. \eqref{hdorig}. In the linearized classical scenario, where the homogeneous background is fixed as a cosmological FLRW solution in GR, the difference between both Hamiltonian functions is just the time derivative of the generating function of the canonical transformation given by Eqs. \eqref{ab} and \eqref{fg}. This change in the Hamiltonian can be realized within the canonical framework of the entire system, that is employed in hybrid quantum cosmology, by completing our change of fermionic variables into a canonical transformation for the full cosmology, that also includes the homogeneous background. This is achieved, at the considered order of perturbative truncation, by correcting the variables of the homogeneous sector with the addition of very specific terms that are quadratic in the fermionic perturbations. In particular, we denote by $a$ the resulting new scale factor. We refer the reader to Refs. \cite{hybr-ref,fermihlqc,backreaction,diagfermi} for the specific details of this procedure. Expressing the total Hamiltonian in terms of the new canonical set of variables for the complete system gives rise to the following contribution \cite{diagfermi}:
\begin{align}\label{newh}
H_D=N_0\sum_{\vec{k},\lambda}\Bigg[2h_D^{k,\lambda}\left(\bar{a}_{\vec{k},\lambda}a_{\vec{k},\lambda}+\bar{b}_{\vec{k},\lambda}b_{\vec{k},\lambda}\right)
+\left\{J_{k,\lambda},H_{|0}\right\}\bar{b}_{\vec{k},\lambda}b_{\vec{k},\lambda} +\bar{h}_I^{k,\lambda}a_{\vec{k},\lambda}b_{-\vec{k},\lambda}-h_I^{\vec{k},(x,y)}\bar{a}_{\vec{k},\lambda}\bar{b}_{-\vec{k},\lambda}\Bigg],
\end{align}
where we have imposed normal ordering of the Grassmann variables, and
\begin{align}\label{hd}
&h_D^{k,\lambda}=\frac{\omega_k}{2a}\left(\left\vert f_2^{k,\lambda}\right\vert ^2-\left\vert f_1^{k,\lambda} \right\vert ^2\right) + M\text{Re}\left(f_1^{k,\lambda}\bar{f}^{k,\lambda}_2\right)+\frac{i}{2}\left(\bar{f}^{k,\lambda}_1 \left\{ f_1^{k,\lambda},H_{|0}\right\}+\bar{f}^{k,\lambda}_2\left\{f_2^{k,\lambda},H_{|0}\right\}\right), \\\label{hi}
&h_I^{k,\lambda}=e^{-iJ_{k,\lambda}}\bigg[if_1^{k,\lambda} \left\{f_2^{k,\lambda},H_{|0}\right\}-if_2^{k,\lambda} \left\{f_1^{k,\lambda},H_{|0}\right\}+2\omega_k a^{-1}f_1^{k,\lambda}f_2^{k,\lambda}+M\left(f_1^{k,\lambda}\right)^2- M\left(f_2^{k,\lambda}\right)^2\bigg].
\end{align}
In all of these expressions, $\left\{.\, ,H_{|0}\right\}$ denotes the Poisson bracket with the Hamiltonian constraint $H_{|0}$ of the homogeneous FLRW inflationary cosmology, evaluated at the new background variables. For concreteness, let us note that this is the cosmological model that can be obtained from ours by ignoring or eliminating the perturbations. Its Hamiltonian is given by $N_0 H_{|0}$. Hence, the considered Poisson bracket is just the derivative with respect to the proper time in the context of linearized classical cosmology. In addition, all the functions of the background in Eq. \eqref{newh} are functionally evaluated on the new canonical variables for the description of this homogeneous sector. The symbol $\text{Re}(\cdot)$ stands for the real part. Finally, in what follows we will restrict $J_{k,\lambda}$ to be constant, in order not to introduce any artificial asymmetry in the dynamics of the annihilation and creationlike variables for particles and antiparticles [see Eq. \eqref{newh}]. 

\subsection{Asymptotic diagonalization}

A look at the Hamiltonian \eqref{newh} immediately shows that, dynamically, the creation and annihilation of pairs of particles and antiparticles is ruled by the function $h_{I}^{k,\lambda}$ of the homogeneous background. Importantly, it is this self-interactive part of the fermionic Hamiltonian what can produce the most severe QFT-type of divergences in the quantum theory \cite{fermihlqc,backreaction}. The issue is directly related to the asymptotic behavior of this function, in the ultraviolet limit of large wavenumbers $\omega_k$. Remarkably, this ultraviolet behavior is greatly tamed by the criteria put forward in hybrid quantum cosmology of (i) requiring that the fermionic annihilation and creation operators can evolve unitarily in the context of QFT in curved spacetimes \cite{uniqfermi,fermihlqc}; and (ii) asking that the Fock representation of the fermionic contribution to the Hamiltonian constraint \eqref{newh} is well defined on the vacuum, something that actually guarantees that certain backreaction effects in the hybrid quantum theory are non-divergent without the need of any regularization \cite{backreaction}. Technically, these criteria succeed in eliminating the first few dominant asymptotic contributions of $h_{I}^{k,\lambda}$ (in powers of the scale $\omega_k$). Furthermore, by imposing these criteria one derives the additional benefit of restricting the asymptotic form of the annihilation and creationlike variables \eqref{ab}, and thus the choice of their Fock representation.

Motivated by these results, a more restrictive criterion, intended for the complete determination of the Fock quantization of the fermionic perturbations, has been recently proposed \cite{diagfermi}. It aims to diminish as much as possible the interaction terms in their Hamiltonian, in the ultraviolet regime of large $\omega_k$. For that, one starts with annihilation and creationlike variables that admit a unitarily implementable dynamics in QFT, within the unique family of unitarily equivalent representations that possess such property, adhering to a standard convention for particles and antiparticles \cite{uniqfermi}. Then, an iterative procedure (that we call asymptotic diagonalization), applied order by order in inverse powers of the Fourier scale $\omega_k$, univocally leads to a complete asymptotic elimination of the interaction terms $h_{I}^{k,\lambda}$, requiring that \cite{diagfermi}
\begin{align}\label{asympdiag}
f_{1}^{k,\lambda}=f_{2}^{k,\lambda}\varphi_{k,\lambda},\qquad\varphi_{k,\lambda}\sim\frac{1}{2\omega_k}\sum_{n=0}^{\infty}\left(-\frac{i}{2\omega_k}\right)^{n}\gamma_{n},
\end{align}
where
\begin{align}\label{recurrence}
\gamma_{0}=Ma,\qquad \gamma_{n+1}=a\left\{H_{|0},\gamma_n\right\}+Ma\sum_{m=0}^{n-1}\gamma_{m}\gamma_{n-(m+1)},\qquad \forall n\geq 0.
\end{align}
We note that, up to the phases $F_{2}^{k,\lambda}$ and $J_{k,\lambda}$, relations \eqref{fg}, \eqref{asympdiag}, and \eqref{recurrence} uniquely provide asymptotic expansions of the coefficients that define the annihilation and creationlike variables for the fermionic perturbations. Specifically, relations \eqref{fg} directly imply that
\begin{align}\label{fg2}
\left\vert f_{2}^{k,\lambda}\right\vert ^2=\frac{1}{1+\left\vert\varphi_{k,\lambda}\right\vert ^2},
\end{align}
so, as an asymptotic expansion, $f_{2}^{k,\lambda}\neq 0$ in our characterization.

The actual hope in hybrid quantum cosmology is that, even if in the form of as an asymptotic expansion, the diagonalization explained above can be used to select (up to the mentioned phases) a complete set of fermionic annihilation and creation operators for all wavevectors $\vec{k}\neq 0$, and hence a specific Fock quantization of the fermionic excitations. In fact, assuming that $f_{2}^{k,\lambda}\neq 0$, as it is required by the standard convention of particles and antiparticles \cite{uniqfermi}, the elimination of interaction terms $h_{I}^{k,\lambda}$ in the fermionic Hamiltonian for all $\omega_k \neq 0$ is attained if and only if
\begin{align}\label{diagallk}
a\left\{\varphi_{k,\lambda},H_{|0}\right\}+2i\omega_k \varphi_{k,\lambda}+iaM\varphi_{k,\lambda}^{2}-iaM=0.
\end{align}
This is a semilinear partial differential equation for $\varphi_{k,\lambda}$, and the current concern regarding the asymptotic diagonalization criterion in hybrid quantum cosmology is whether the asymptotic expansion \eqref{asympdiag} uniquely characterizes a solution. In the upcoming section, we argue that this is indeed the case when the homogeneous background describes a de Sitter spacetime, and the fermionic perturbations are considered within the linearized context of QFT in curved spacetimes.

\section{Unique vacuum in de Sitter}\label{desitter}

We now restrict all our attention to the scenario of cosmological models with negligible backreaction of the perturbations on the homogeneous background, situation in which this background follows the classical dynamics of an FLRW spacetime fuelled with a homogeneous scalar field. In practice, this means that the old and new variables that describe the homogeneous cosmology can be identified (in particular, we have $a=\tilde{a}$), and that the Poisson bracket $a\left\{.\, ,H_{|0}\right\}$ is the derivative with respect to conformal time, $\eta$. The metric that describes the de Sitter spacetime is a particular solution of the considered, classical flat FLRW cosmologies, expressed in coordinates that correspond to its flat slicing \cite{mukh}. Specifically, this solution can be reached with a constant potential for the scalar field. In conformal time, the scale factor then behaves as
\begin{align}\label{scaledsitter}
a=-(\eta H_{\Lambda})^{-1}, \qquad -\infty<\eta<0,
\end{align}
where $H_{\Lambda}$ is the Hubble constant. In this de Sitter background, the general condition \eqref{diagallk} that cancels the interaction terms in the fermionic Hamiltonian becomes the following Riccati equation
\begin{align}\label{eqdesitter}
\varphi_{k,\lambda}^{\prime}+2i\omega_k \varphi_{k,\lambda} -iM(\eta H_{\Lambda})^{-1}\varphi_{k,\lambda}^{2}+iM(\eta H_{\Lambda})^{-1}=0,
\end{align}
where the prime denotes the derivative with respect to $\eta$. In order to eventually find the general solution to this equation, we introduce the standard change of variable
\begin{align}
\varphi_{k,\lambda}=i\eta M^{-1}H_{\Lambda}(\log{u_{k,\lambda}})^{\prime},
\end{align}
which leads to the second-order linear equation
\begin{align}
u_{k,\lambda}^{\prime\prime}+\left(2i\omega_k+ \eta^{-1}\right)u_{k,\lambda}^{\prime}+\left(M^{-1}H_{\Lambda}\eta\right)^{-2}u_{k,\lambda}=0.
\end{align}
We can bring this equation to a  well-known ordinary differential equation if we redefine $u_{k,\lambda}=e^{iMt}v_{k,\lambda}$, where $t$ is the comoving cosmological time, in terms of which the scale factor is $a= \exp{(H_{\Lambda}t)}$. In this way, and introducing the mode-dependent complex time $T_k=-2i\omega_k\eta$, we finally arrive at
\begin{align}\label{confluent}
T_k\frac{d^2 v_{k,\lambda}}{dT_k^2}+\left(1-2iMH_{\Lambda}^{-1}-T_k\right)\frac{d v_{k,\lambda}}{dT_k}+iMH_{\Lambda}^{-1}v_{k,\lambda}=0.
\end{align}
This is a confluent hypergeometric equation in the complex variable $T_k$ \cite{spfunc}. Its general solution is given by the following linear combination of convergent hypergeometric functions:
\begin{align}\label{generalv}
v_{k,\lambda}=A\,{}_{1}F_{1}\left(-iMH_{\Lambda}^{-1};1-2iMH_{\Lambda}^{-1};T_k\right)+B\,T_k^{2iMH_{\Lambda}^{-1}}{}_{1}F_{1}\left(iMH_{\Lambda}^{-1};1+2iMH_{\Lambda}^{-1};T_k\right),
\end{align}
where $A$ and $B$ are arbitrary complex integration constants that may in general depend on $\omega_k$ and $\lambda$, even if we have not indicated explicitly this possibility. For concreteness, we recall the definition of the hypergeometric function of type $(p,q)$, as a formal power series,
\begin{align}
&{}_{p}F_{q}\left(b_{1},...,b_{p};c_{1},...,c_{q};z\right)=\sum_{n=0}^{\infty}\frac{(b_1)_n ... (b_p)_n}{(c_1)_n ... (c_q)_n}\frac{z^n}{n!},\quad 
&(b)_{n}=\begin{cases}
1 & \text{if }n=0, \\
b(b+1)...(b+n-1) & \text{if }n>0,
\end{cases}\label{CASES}
\end{align}
for $b$ equal to any of the complex numbers $b_{1},...,b_{p},c_{1},...,c_{q}$. Let us point out that this series converges absolutely for all $z$ if $p\leq q$, while it has a vanishing radius of convergence if $p>q+1$ \cite{spfunc}.

\subsection{Uniqueness from asymptotic diagonalization}

In the de Sitter background, formula \eqref{generalv} can be used to obtain the form of the general solution $\varphi_{k,\lambda}=f_1^{k,\lambda}(f_2^{k,\lambda})^{-1}$ to Eq. \eqref{eqdesitter} and, by means of relation \eqref{fg2}, the coefficients for the definition of fermionic annihilation and creationlike variables that display no dynamical self-interaction. In what follows, we show that the criterion of asymptotic diagonalization, that leads to the asymptotic expansion given in Eqs. \eqref{asympdiag} and \eqref{recurrence}, uniquely determines a pair of integration constants $A$ and $B$ in Eq. \eqref{generalv} (up to a global multiplicative factor), and hence a unique solution $\varphi_{k,\lambda}$.

Let us start by studying the iterative relation \eqref{recurrence} for the coefficients $\gamma_n$ that appear in the asymptotic diagonalization expansion, for the classical de Sitter cosmological background. In the considered linearized context, it reads
\begin{align}
\gamma_{0}=-M(H_{\Lambda}\eta)^{-1},\qquad \gamma_{n+1}=-\gamma_n^{\prime}-M(H_{\Lambda}\eta)^{-1}\sum_{m=0}^{n-1}\gamma_{m}\gamma_{n-(m+1)},\qquad \forall n\geq 0.
\end{align}
It is not hard to check that its solution leads to an asymptotic expansion for $\varphi_{k,\lambda}$ of the form
\begin{align}\label{asympphi}
&\varphi_{k,\lambda}\sim iT_k^{-1}\sum_{n=0}^{\infty}(-T_k)^{-n}C_{n}, \nonumber \\  & C_0=MH_{\Lambda}^{-1},\qquad C_{n+1}=(n+1)C_n +MH_{\Lambda}^{-1}\sum_{m=0}^{n-1}C_{m}C_{n-(m+1)},\qquad \forall n\geq 0.
\end{align}
We do not need to solve the complicated iterative equation for these coefficients, as all the relevant information is contained in the associated expansion of $v_{k,\lambda}$, that we explicitly determine below. In fact, the deduced expression greatly constrains the asymptotic behavior of the corresponding, particular solution $v_{k,\lambda}$ of the confluent hypergeometric equation \eqref{confluent}. Indeed, since
\begin{align}\label{varphiv}
\varphi_{k,\lambda}=1+iM^{-1}H_{\Lambda}T_k\frac{d}{dT_k}\left(\log{v_{k,\lambda}}\right),
\end{align}
the asymptotic expansion in inverse powers of $T_k$ that we have obtained for $\varphi_{k,\lambda}$ implies that, necessarily,
\begin{align}
v_{k,\lambda}\sim T_{k}^{iMH_{\Lambda}^{-1}}\sum_{n=0}^{\infty}(-T_k)^{-n}v_{n},\qquad \text{with}\qquad v_1=\left(MH_{\Lambda}^{-1}\right)^2 v_0.
\end{align}
The imaginary power of $T_k$ that appears in the above expression in fact is needed to eliminate the term of order $1$ in $\varphi_{k,\lambda}$, so that the function $T_k \varphi_{k,\lambda}$ dominantly behaves like $iMH_{\Lambda}^{-1}$ when $\omega_k\rightarrow\infty$, as it is required by Eq. \eqref{asympphi}. If we introduce this asymptotic expansion for $v_{k,\lambda}$ in the confluent hypergeometric equation that it must satisfy, we find a recursion relation for its constant coefficients $v_{n}$,
\begin{align}
v_{n+1}=\frac{\left(n+iMH_{\Lambda}^{-1}\right)\left(n-iMH_{\Lambda}^{-1}\right)}{n+1}v_n.
\end{align}
The solution is clearly
\begin{align}
v_n=\frac{v_0}{n!}\left(iMH_{\Lambda}^{-1}\right)_{n}\left(-iMH_{\Lambda}^{-1}\right)_{n}
\end{align}
for an arbitrary constant $v_0$, and where we have used the notation introduced in Eq. \eqref{CASES}. So, the asymptotic expansion selected for $v_{k,\lambda}$ by our Hamiltonian diagonalization corresponds to the hypergeometric function
\begin{align}\label{asympv}
v_{k,\lambda}\sim v_{0}T_{k}^{iMH_{\Lambda}^{-1}}{}_{2}F_{0}\left(iMH_{\Lambda}^{-1},-iMH_{\Lambda}^{-1};-;-T_k^{-1}\right),
\end{align}
that has a vanishing radius of convergence. Here, the hyphen between semicolons in the argument of the hypergeometric function just indicates the case $q=0$ of its definition, case for which the denominator in Eq. \eqref{CASES} becomes the factorial of $n$. Even though it is formally divergent, this is precisely the asymptotic expansion (up to the global factor $v_0$) of a very particular recessive solution of the confluent equation, known as the Tricomi solution \cite{spfunc,abram}. We now explicitly prove that this solution is actually the only one that admits such an asymptotic behavior. To do so, we first need the asymptotic expansion of the general solution \eqref{generalv} for arbitrary constants $A$ and $B$. Actually, for $-\pi/2<\text{arg}(z)<3\pi/2$, it holds that \cite{abram}
\begin{align}\label{asym1f1}
{}_{1}F_{1} \left(b;c;z\right)\sim \frac{\Gamma(c)}{\Gamma(c-b)}z^{-b}e^{i\pi b}{}_{2}F_{0}\left(b,1+b-c;-;-z^{-1}\right)+\frac{\Gamma(c)}{\Gamma(b)}z^{b-c}e^{z}{}_{2}F_{0}\left(c-b,1-b;-;z^{-1}\right),
\end{align}
so the general solution \eqref{generalv} of our confluent equation has the following asymptotic expansion with respect to $T_k$:
\begin{align}\label{asympgeneralv}
v_{k,\lambda} \sim &T_k^{iMH_{\Lambda}^{-1}}{}_{2}F_{0}\left(iMH_{\Lambda}^{-1},-iMH_{\Lambda}^{-1};-;-T_k^{-1}\right)\left[A\frac{\Gamma\left(1-2iMH_{\Lambda}^{-1}\right)}{\Gamma\left(1-iMH_{\Lambda}^{-1}\right)}e^{\pi MH_{\Lambda}^{-1}}+B\frac{\Gamma\left(1+2iMH_{\Lambda}^{-1}\right)}{\Gamma\left(1+iMH_{\Lambda}^{-1}\right)}e^{-\pi MH_{\Lambda}^{-1}}\right]\\ \nonumber &+e^{T_k}T_k^{-1+iMH_{\Lambda}^{-1}}{}_{2}F_{0}\left(1-iMH_{\Lambda}^{-1},1+iMH_{\Lambda}^{-1};-;T_k^{-1}\right) \left[A\frac{\Gamma\left(1-2iMH_{\Lambda}^{-1}\right)}{\Gamma\left(-iMH_{\Lambda}^{-1}\right)}+B\frac{\Gamma\left(1+2iMH_{\Lambda}^{-1}\right)}{\Gamma\left(iMH_{\Lambda}^{-1}\right)}\right].
\end{align}
The two terms on the right-hand side of this expression clearly represent (even if only formally) two linearly independent functions of $T_k$, whereas just one of them appears in the expansion \eqref{asympv} that is selected by the asymptotic diagonalization criterion. Therefore, a necessary condition imposed by this criterion is that
\begin{align}
B=-\frac{\Gamma\left(iMH_{\Lambda}^{-1}\right)\Gamma\left(1-2iMH_{\Lambda}^{-1}\right)}{\Gamma\left(1+2iMH_{\Lambda}^{-1}\right)\Gamma\left(-iMH_{\Lambda}^{-1}\right)}A.
\end{align}
Introducing this value of $B$ in the general formula \eqref{asympgeneralv}, and using the general property of the Gamma function $\Gamma(1+z)=z\Gamma(z)$, with $z\in\mathbb{C}$ \cite{abram}, we obtain
\begin{align}
v_{k,\lambda} \sim AT_k^{iMH_{\Lambda}^{-1}}4\cosh{\left(\pi MH_{\Lambda}^{-1}\right)}\frac{\Gamma\left(-2iMH_{\Lambda}^{-1}\right)}{\Gamma\left(-iMH_{\Lambda}^{-1}\right)}{}_{2}F_{0}\left(iMH_{\Lambda}^{-1},-iMH_{\Lambda}^{-1};-;-T_k^{-1}\right).
\end{align}
Comparing once again with the asymptotic expansion \eqref{asympv} of our desired solution, we can determine the value of $A$. In this way, we are univocally led to conclude that
\begin{align}\label{ABcoefs}
A=v_{0}\frac{\Gamma\left(2iMH_{\Lambda}^{-1}\right)}{\Gamma\left(iMH_{\Lambda}^{-1}\right)},\quad B=v_0\frac{\Gamma\left(-2iMH_{\Lambda}^{-1}\right)}{\Gamma\left(-iMH_{\Lambda}^{-1}\right)},
\end{align}
where we have employed the general identity \cite{abram}
\begin{align}\nonumber
\frac{1}{4\cosh{(\pi y)}}=\frac{\Gamma(2iy)\Gamma(-2iy)}{\Gamma(iy)\Gamma(-iy)},\qquad y\in\mathbb{R}.
\end{align}

We have henceforth proven that, in a de Sitter background, our asymptotic characterization inspired by hybrid quantum cosmology uniquely picks out a particular solution of the confluent hypergeometric equation \eqref{confluent} (up to the irrelevant factor $v_0$), and therefore a particular function $\varphi_{k,\lambda}$ [cf. Eq. \eqref{varphiv}] that eliminates the self-interaction in the fermionic Hamiltonian for all $\vec{k}\neq0$. The specification of $\varphi_{k,\lambda}$, in turn, corresponds to a precise choice of the fermionic annihilation and creationlike variables \eqref{ab}, up to the two phases $F_{2}^{k,\lambda}$ and $J_{k,\lambda}$, and thus to a unique Fock representation (with its associated vacuum state) of the Dirac field. In particular, the selected solution $v_{k,\lambda}$ is given by Eq. \eqref{generalv} after substituting the constant coefficients $A$ and $B$ by the values given in Eq. \eqref{ABcoefs}. Up to the constant factor $v_0$, the result is then the recessive solution of the confluent hypergeometric equation commonly known as the Tricomi function, usually expressed as $U(-iMH_{\Lambda}^{-1},1-2iMH_{\Lambda}^{-1},T_k)$ \cite{spfunc,abram}.

\subsection{Field decomposition}

In this final subsection we explicitly compute the basis of solutions of the Dirac equation in de Sitter cosmology that is associated with the choice of Fock representation selected by the asymptotic diagonalization criterion. It is in terms of this basis that the quantum representation of the Dirac field, viewed as an operator valued distribution, can be decomposed, and the coefficients in such decomposition are the annihilation and creation operators for particles and antiparticles. The fermionic vacuum state in the resulting Fock space is then uniquely specified (up to a phase) as the state that vanishes upon the action of all of the annihilation operators. 

In order to obtain this field decomposition, let us first notice that, combining Eqs. \eqref{mdecomp}, \eqref{ab}, and \eqref{fg}, one can express the Dirac field in terms of any canonical set of annihilation and creationlike variables. In the context of QFT in classical cosmological spacetimes, these variables obey the dynamics dictated by the Hamiltonian \eqref{newh}-\eqref{hi} (where the Poisson brackets must be replaced with the corresponding time derivatives). Then, for variables that display no dynamical self-interaction, namely for coefficients $f_{1}^{k,\lambda}$ and $f_{2}^{k,\lambda}$ such that Eq. \eqref{diagallk} holds, we can write the Dirac field as
\begin{align}\label{fdecomp}
\psi(\eta,\vec{x})=\sum_{\vec{k},\lambda}
\left[u_{\vec{k},\lambda}(\eta,\vec{x})A_{\vec{k},\lambda} + w_{\vec{k},\lambda}(\eta,\vec{x})\bar{B}_{\vec{k},\lambda} \right],
\end{align}
where $A_{\vec{k},\lambda}=a_{\vec{k},\lambda}(\eta_0)$ and $B_{\vec{k},\lambda}=b_{\vec{k},\lambda}(\eta_0)$ are the constant annihilation coefficients for particles and antiparticles (to be promoted to the corresponding operators in the Schr\"odinger picture) and $\eta_0$ is an arbitrary choice of initial time employed for their definition. In addition, the basis elements are
\begin{align}\label{uv1}
&u_{\vec{k},\lambda}(\eta,\vec{x})=\frac{e^{i2\pi \vec{k}\vec{x}/l_0}}{\sqrt{ l_0^3 a^3}}\left[I-\frac{1-\lambda}{2}\left(I+i\gamma^{0}\right)\right]\bar{f}_{2}^{k,\lambda}e^{-i\Omega_k(\eta,\eta_0)}\begin{pmatrix}
\bar{\varphi}_{k,\lambda}(\eta)\xi_{\lambda}(\vec{k}) \\
\xi_{\lambda}(\vec{k})\end{pmatrix}, \\ \label{uv2} & w_{\vec{k},\lambda}(\eta,\vec{x})=-e^{-iJ_{k,\lambda}(\eta_0)}\lambda\gamma^2 \bar{u}_{\vec{k},\lambda}(\eta,\vec{x}),
\end{align}
where the integrated time-dependent ``frequency'' of the diagonal evolution of the annihilationlike variables is
\begin{align}
\Omega_k(\eta,\eta_0)=2\int_{\eta_0}^{\eta} d\tilde\eta\, a(\tilde\eta)h_{D}^{k,\lambda}(\tilde\eta).
\end{align}
After using the partial differential equation \eqref{diagallk} and the relation \eqref{fg2}, the canonical expression for this frequency is found to be given by
\begin{align}\label{hddiag}
2h_{D}^{k,\lambda}=a^{-1}\omega_k +M\text{Re}\left(\varphi_{k,\lambda}\right)-\left\{F_{2}^{k,\lambda},H_{|0}\right\}.
\end{align}
It is worth noticing that this formula would hold as well in the full context of hybrid quantum cosmology, employing the perturbatively corrected variables for the homogeneous cosmological sector, once a solution of equation \eqref{diagallk} had been constructed (ideally by following the criterion of asymptotic diagonalization).

In the de Sitter cosmology under analysis, we recall that the function $\varphi_{k,\lambda}$ is obtained from the solution $v_{k,\lambda}$ of the confluent equation by means of Eq. \eqref{varphiv}, that involves the logarithmic derivative of this solution with respect to the imaginary time $T_k=-2i\omega_k\eta$. We have proven that the condition of asymptotic diagonalization, inspired by hybrid quantum cosmology, serves to select a unique $v_{k,\lambda}$, given by the Tricomi function $U\left(-iMH_{\Lambda}^{-1},1-2iMH_{\Lambda}^{-1},T_k\right)$ multiplied by a constant factor $v_0$. The derivatives of this function have been studied in detail \cite{abram}, and in our case we have
\begin{align}
\frac{dv_{k,\lambda}}{dT_k}=iv_0 MH_{\Lambda}^{-1} U\left(1-iMH_{\Lambda}^{-1},2-2iMH_{\Lambda}^{-1},T_k\right).
\end{align}
This is a Tricomi function of the form $U(\bar\mu+1/2,2\bar\mu+1,-2iz)$, with $\mu=iMH_{\Lambda}^{-1}+1/2$ and $z=\omega_k\eta$. Tricomi functions of this special type satisfy the identity \cite{abram}
\begin{align}\label{specialU}
U\left(\nu+\frac{1}{2},2\nu+1,-2iz\right)=\frac{\sqrt{\pi}}{2}ie^{i(\pi\nu-z)}(2z)^{-\nu}H_{\nu}^{(1)}(z),
\end{align}
that relates them to the Hankel function of the first kind $H_{\nu}^{(1)}$. Applying this property to our solution, we get
\begin{align}\label{vderiv}
\frac{dv_{k,\lambda}}{dT_k}=-iv_{0}\frac{\sqrt{\pi}}{2}MH_{\Lambda}^{-1}e^{MH_{\Lambda}^{-1}}e^{-i\omega_k\eta}(2\omega_k\eta)^{\mu -1}H^{(1)}_{1-\mu}(\omega_k\eta).
\end{align}
On the other hand, the Tricomi function that appears as the denominator of the logarithmic derivative of $v_{k,\lambda}$ can also be expressed in terms of Hankel functions by using the recursive relation \cite{abram}
\begin{align}
U\left(-iMH_{\Lambda}^{-1},1-2iMH_{\Lambda}^{-1},T_k\right)=\frac{1}{2}U\left(-iMH_{\Lambda}^{-1},-2iMH_{\Lambda}^{-1},T_k\right)+\frac{T_k}{2}U\left(1-iMH_{\Lambda}^{-1},2-2iMH_{\Lambda}^{-1},T_k\right).
\end{align}
The two functions on the right-hand side are of the special form \eqref{specialU}, and hence we can write
\begin{align}\label{vdenom}
\frac{v_{k,\lambda}(\eta)}{v_0}=\frac{\sqrt{\pi}}{4}e^{MH_{\Lambda}^{-1}}e^{-i\omega_k\eta}(2\omega_k\eta)^{\mu}\left[H^{(1)}_{-\mu}(\omega_k\eta)+iH^{(1)}_{1-\mu}(\omega_k\eta)\right].
\end{align}
Introducing expressions \eqref{vderiv} and \eqref{vdenom} in the relation \eqref{varphiv} between $v_{k,\lambda}$ and $\varphi_{k,\lambda}$, we find the explicit form of this function selected by the asymptotic diagonalization criterion in de Sitter,
\begin{align}\label{varphidesitter}
\varphi_{k,\lambda}(\eta)=\frac{H^{(1)}_{-\mu}(\omega_k\eta)-iH^{(1)}_{1-\mu}(\omega_k\eta)}{H^{(1)}_{-\mu}(\omega_k\eta)+iH^{(1)}_{1-\mu}(\omega_k\eta)}.
\end{align}
Its complex conjugate, that directly appears in the basis decomposition \eqref{fdecomp}-\eqref{uv2}, is then simply
\begin{align}\label{varphiconj}
\bar\varphi_{k,\lambda}(\eta)=\frac{H^{(2)}_{\mu -1}(\omega_k\eta)+iH^{(2)}_{\mu}(\omega_k\eta)}{H^{(2)}_{\mu -1}(\omega_k\eta)-iH^{(2)}_{\mu}(\omega_k\eta)},
\end{align}
where $H^{(2)}_{\nu}$ is the Hankel function of the second kind and we have used that $\overline{{H}^{(1)}_{\nu}(z)}=H^{(2)}_{\bar\nu}(\bar{z})$ \cite{abram}. This function $\varphi_{k,\lambda}$, in turn, contains all the information about the norm of $f_{2}^{k,\lambda}$ as displayed in relation \eqref{fg2}, that encodes the canonical anticommutation algebra of the annihilation and creationlike variables. Explicitly, we obtain the result
\begin{align}\label{f2}
\left\vert f_{2}^{k,\lambda}\right\vert ^2=\frac{\pi\omega_k\eta}{8}e^{\pi MH_{\Lambda}^{-1}}\left\vert H^{(2)}_{\mu -1}(\omega_k\eta)-iH^{(2)}_{\mu}(\omega_k\eta)\right\vert ^2,
\end{align}
after some algebraic manipulations and using the following identity for the Wronskian of Hankel functions \cite{abram}:
\begin{align}\nonumber
H_{1-\nu}^{(1)}(z)H_{\nu}^{(2)}(z)+H_{-\nu}^{(1)}(z)H_{\nu -1}^{(2)}(z)=-\frac{4i}{\pi z}e^{i\pi\nu}.
\end{align}

The only quantity that remains to be determined in order to reach the final form of the decomposition \eqref{fdecomp}-\eqref{uv2} of the Dirac field selected by our criterion is the time dependent frequency $\Omega_k(\eta,\eta_0)$. If we particularize the formula for the coefficients $2h_{D}^{k,\lambda}$ of the diagonal fermionic Hamiltonian, given in Eq. \eqref{hddiag}, to the considered case of a homogeneous de Sitter background with no backreaction, and the function $\varphi_{k,\lambda}$ is identified as the specific one singled out by our criterion, we get that, up to an additive constant,
\begin{align}\label{omegadesitter}
\Omega_k=\theta_k-F_{2}^{k,\lambda},\qquad \theta_k (\eta)=\arg{\left[H^{(1)}_{-\mu}(\omega_k\eta)+iH^{(1)}_{1-\mu}(\omega_k\eta)\right]},
\end{align}
where $\arg{[.]}$ denotes the phase of its complex argument. Combining all our results [cf. Eqs. \eqref{varphiconj}, \eqref{f2}, and \eqref{omegadesitter}], the basis of solutions that describes the particles associated with the Fock representation of the Dirac field in de Sitter, uniquely picked out by the asymptotic diagonalization criterion, reads
\begin{align}\label{finalu}
u_{\vec{k},\lambda}(\eta,\vec{x})=\frac{e^{i2\pi \vec{k}\vec{x}/l_0}}{\sqrt{ l_0^3 a^3}}\left[I-\frac{1-\lambda}{2}\left(I+i\gamma^{0}\right)\right]\sqrt{\frac{\pi\omega_k \eta}{8}}e^{i\Theta+\pi MH_{\Lambda}^{-1}/2}\begin{pmatrix}
[H^{(2)}_{\mu -1}(\omega_k\eta)+iH^{(2)}_{\mu}(\omega_k\eta)]\xi_{\lambda}(\vec{k}) \\
[H^{(2)}_{\mu -1}(\omega_k\eta)-iH^{(2)}_{\mu}(\omega_k\eta)]\xi_{\lambda}(\vec{k})\end{pmatrix},
\end{align}
where $\Theta$ is a constant global phase, that is irrelevant for the definition of the vacuum. On the other hand, the solutions that describe antiparticles are given by the charge conjugate of these ones, namely via Eq. \eqref{uv2}. It is worth noticing that the constant phase $\Theta$ includes all possible dependence of the basis of solutions on the choice of initial time $\eta_0$ for the definition of the annihilation and creationlike constant coefficients, and thus the vacuum that results from our approach is independent of that choice. We also note that, in the asymptotic regime of large $k$, the leading time dependence of our basis of solutions determined by $u_{\vec{k},\lambda}$ follows the behavior $\exp{(-i\omega_k \eta)}$, up to multiplication by $a^{-3/2}$ and a constant, something that is often required as a necessary physical feature of the corresponding Fock representation of fields in conformally flat spacetimes \cite{bdf2,radz,hanno}. In particular, the BD Hadamard vacuum for scalar fields in de Sitter displays such a dominant plane wave behavior \cite{mukh}.

In order to establish more precisely the connection between our result and the statements available in the literature about the choice of vacuum state for Dirac fields in de Sitter, we end this subsection by discussing the relation between the mode decomposition of our solutions and the one assigned in Ref. \cite{bdf2} as corresponding to the BD state. First of all, for such a comparison we need to change from the Weyl representation of the constant Clifford algebra (employed here) to the Dirac representation (used in Ref. \cite{bdf2}). They are related by a unitary change of the spinorial basis for the Dirac field, namely
\begin{align}
\psi^{W}=T\psi^{D},\qquad T=\frac{1}{\sqrt{2}}\begin{pmatrix}I &-I \\ I & I\end{pmatrix},
\end{align}
where the superscripts $W$ and $D$ indicate objects in the Weyl and Dirac representations, respectively.  For our mode decomposition, this change leads to a basis in the Dirac representation given (up to a global constant phase) by
\begin{align}\label{udirac}
u^{D}_{\vec{k},\lambda}(\eta,\vec{x})=\frac{e^{i2\pi \vec{k}\vec{x}/l_0}}{\sqrt{ l_0^3 a^3}}\frac{\sqrt{\pi\omega_k \eta}}{2}e^{\pi MH_{\Lambda}^{-1}/2}\begin{pmatrix}
i\lambda H^{(2)}_{\mu -1}(\omega_k\eta)\xi_{\lambda}(\vec{k}) \\
 H^{(2)}_{\mu}(\omega_k\eta)\xi_{\lambda}(\vec{k})\end{pmatrix},
\end{align}
together with Eq. \eqref{uv2} for the charge conjugate counterpart. The resulting basis of solutions is exactly the same as that corresponding to the BD state in Ref. \cite{bdf2}, after one interchanges the two (bidimensional) components of the spinor on the right-hand side. Actually, this difference can be attributed just to a change in the global sign of the tetrads that are employed in the two compared works, as we now briefly explain. With our $(-+++)$ convention for the Minkowski metric, the Dirac equation is
\begin{align}
e^{\nu}_{b}\gamma^{b}D_{\nu}\psi - M\psi=0, 
\end{align}
where $\gamma^{b}$ are the generators of the constant Clifford algebra in any representation, and $D_{\nu}$ is the spin covariant derivative \cite{SGeom}. Here, $\nu$ denotes a spacetime tensor index, whereas $b=0,... 3$ is an internal Lorentz index. The spin covariant derivative contains a connection one-form which depends on the tetrads $e^{\nu}_{b}$ and their derivatives only through quadratic and quartic products. It then follows that interchanging two choices of tetrad that differ only in a global multiplicative sign has exclusively the net effect of an apparent flip of sign in the mass term of the Dirac equation, in what concerns the choice of gauge for the Dirac field as a solution to this equation. According to our comments in Sec. \ref{diaggeneral}, in the expanding flat chart of de Sitter (in conformal time) we have selected the tetrad as $e^{\nu}_{b}=a^{-1}\delta^{\nu}_{b}$, where the scale factor $a$ is given in Eq. \eqref{scaledsitter}. The choice employed in Ref. \cite{bdf2} is precisely the opposite in sign, i.e., it is given by minus this tetrad. For both choices, if one then works, e.g., in the Dirac representation of the Clifford algebra, the Dirac equation can be recast as a second order Bessel differential equation in $x=\omega_k \eta$ for the time-dependent factor of the first (bidimensional) component of $\eta^{-2} u^{D}_{\vec{k},\lambda}$. The second component is completely fixed in terms of the first one by means of the Dirac equation. The difference between the two considered conventions in the choice of tetrad (ours and that of Ref. \cite{bdf2}) is reflected in the order of the Bessel equation, that becomes, respectively, $\mu -1$ and $\mu$. The solutions that have an asymptotic behavior with a dominant time-dependence proportional to $x^{-1/2}\exp{(-ix)}$ are uniquely given, respectively, by $H^{(2)}_{\mu -1}(x)$ and $H^{(2)}_{\mu}(x)$ \cite{abram}. The second component of $\eta^{-2}u^{D}_{\vec{k},\lambda}$ then results proportional to  $H^{(2)}_{\mu}(x)$ and $H^{(2)}_{\mu -1}(x)$, respectively. The remaining factors of the spinor \eqref{udirac} are determined by the normalization of the solutions \cite{bdf2}. We recall that the above Hankel functions are precisely the two parts that must be interchanged in Eq. \eqref{udirac} in order to identify the bases of solutions constructed in the two considered works, up to a global phase. Therefore, we conclude that the vacuum state for the Dirac field resulting from our analysis corresponds indeed to the BD state defined in Ref. \cite{bdf2}, once the same choice of local Lorentz gauge is made.

\section{Conclusions}\label{conclusions}

In this paper we have shown how the criterion of asymptotic diagonalization, originated in the framework of hybrid quantum cosmology, can serve to single out a privileged Fock representation of the Dirac field in de Sitter spacetime, within the context of QFT in curved spacetimes. The explicit basis of solutions to the Dirac equation associated with that choice of representation has also been computed, in terms of Hankel functions of the first and second kind, and in coordinates associated with the conformal flat slicing of de Sitter. Furthermore, the canonical expression for the resulting diagonal Hamiltonian that dictates the linearized dynamics of the annihilation and creationlike variables selected by our criterion has been found, exclusively in terms of them and functions of the homogeneous variables that describe the cosmological background. In particular, the derived formula \eqref{hddiag} could be of potential use in hybrid quantum cosmology, if the asymptotic diagonalization problem is solved for more general cosmological backgrounds than de Sitter.

The hybrid approach to quantum FLRW cosmology with perturbations contemplates the natural freedom of making a dynamical splitting between the spatially homogeneous, global, degrees of freedom of the system and the inhomogeneous perturbations. When these perturbations consist of a Dirac field, such freedom can be encoded in choices of annihilation and creationlike variables, given by linear transformations of the field mode coefficients that depend explicitly on the homogeneous background. These transformations can be completed so that they become canonical for the entire cosmological system, truncated at quadratic perturbative order in the action, a procedure that leads to a fermionic contribution to the total Hamiltonian that dictates the linearized dynamics of the annihilation and creationlike variables. The criterion of asymptotic diagonalization consists in restricting almost all the freedom in the selection of these variables, and with it the aforementioned dynamical splitting, together with the Fock quantization of the fermionic degrees of freedom, so that the fermionic Hamiltonian gets diagonalized in a way that is adapted to its local structure. This strategy provides a very precise asymptotic expansion of the functions of the background that define the annihilation and creationlike variables. In turn, this determines the expansion of at least one solution of the partial differential equation \eqref{diagallk}, that arises from the general demand that the Hamiltonian become diagonal.

If one disregards all backreaction effects of the perturbations on the homogeneous background, and considers that this cosmological background is just a solution of the classical FLRW spacetime, the possible definitions of annihilation and creationlike variables compatible with the requirements of hybrid quantum cosmology can directly be understood as different choices of Fock representations of the Dirac field in the context of QFT in curved spacetimes. We have restricted our attention to this situation and, furthermore, we have particularized the FLRW background, identifying it with a de Sitter solution. Then, we have shown that the asymptotic expansion selected by the criterion of asymptotic diagonalization indeed picks out a {\emph {unique}} function among those that define annihilation and creationlike variables that follow a diagonal evolution for all spatial scales. The basis of solutions to the Dirac equation associated with the resulting Fock representation of the field can be specified completely and in an analytical way.

Our result is potentially relevant within QFT, as well as in the context of quantum cosmology. On the one hand, we have explicitly provided a unique fermionic vacuum in de Sitter spacetime selected by a very well characterized criterion, that has its original motivation in hybrid quantum cosmology. This sheds further light on the question of which is the natural analog of the BD vacuum for a Dirac field. In this context, our analysis precisely leads to the basis of solutions for the field that has been assigned to correspond to such a fermionic BD state in the literature \cite{bdf2}. In particular, it displays the ultraviolet behavior expected to guarantee Hadamard-like properties. This result supports the potential robustness of our criterion to select a privileged fermionic vacuum state in FLRW cosmologies, specially considering that the very same criterion of asymptotic diagonalization has succeeded in predicting the BD vacuum for the well-known cases of scalar and tensor perturbations \cite{diagscalar}. On the other hand, our work shows that, at least in certain cases, the criterion employed in hybrid quantum cosmology to i) determine a dynamical splitting between the homogeneous and inhomogeneous sectors in phase space, and ii) select a Fock representation for the inhomogeneities, can indeed result into a complete removal of both types of ambiguities, even when this criterion was initially based solely on ultraviolet considerations.

\acknowledgements

This work was supported by Project. No. MINECO FIS2017-86497-C2-2-P from Spain.

\end{document}